# A Modern Approach to Electoral Delimitation using the Quadtree Data Structure

Sahil Kale
Computer Engineering
Pune Institute of Computer Technology
Pune, India
sahilrkale05@gmail.com

Gautam Khaire
Computer Engineering
Pune Institute of Computer Technology
Pune, India
gautamkhaire06052002@gmail.com

Jay Patankar
Computer Engineering
Pune Institute of Computer Technology
Pune, India
patankarjays@gmail.com

Pujashree Vidap
Computer Engineering
Pune Institute of Computer Technology
Pune, India
psvidap@pict.edu

*Abstract*—The boundaries of electoral constituencies for assembly and parliamentary seats are drafted using a process referred to as delimitation, which ensures fair and equal representation of all citizens. The current delimitation exercise suffers from a number of drawbacks viz. inefficiency, gerrymandering and an uneven seat to population ratio, owing to existing legal and constitutional dictates. The existing methods allocate seats to every state but remain silent about their actual shape and location within the state. The main purpose of this research is to study and analyse the performance of existing delimitation algorithms and further propose a potential solution, along with its merits, that involves using a computational model based on the quadtree data structure to automate the districting process by optimizing objective population criteria.

The paper presents an approach to electoral delimitation using the quadtree data structure, which is used to partition a two-dimensional geographical space by recursively subdividing it into four quadrants or regions on the basis of population as a parameter value associated with the node. The quadtree makes use of a quadrant schema of the geographical space for representing constituencies, which not only keeps count of the allocated constituencies but also holds their location specific information. The performance of the proposed algorithm is analysed and evaluated against existing techniques and proves to be an efficient solution in terms of algorithmic complexity and boundary visualisation to the process of political districting.

*Keywords—component, formatting, style, styling, insert (key words)*

## I. INTRODUCTION

Constituencies represent the crux of democracy by forming the arenas for representation of citizens by elected candidates from political parties. The process of electoral delimitation refers to the process of demarcating limits or boundaries of territorial constituencies in a country or a province having a legislative body [1-2]. Electoral delimitation plays a significant role in maintaining proportional representation in the electoral system as it governs the proportionality of the number of representative seats allocated to an electoral district and governs the democratic fairness among territories in a region [3].

Delimitation being such a complex and cumbersome exercise, most of the countries in the world entrust the task to separately elected bodies. In the US, Canada, France, Italy, Germany, Belgium, Poland, Switzerland, the legislatures deal with the task of redistricting, while in India, Australia and the UK, a separate Delimitation Commission has been set up to resolve the delimitation process [2].

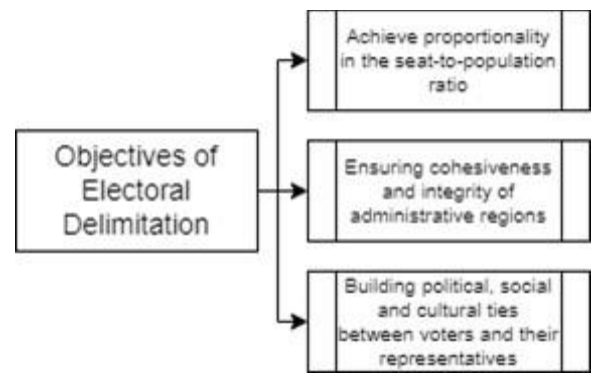

Fig. 1. Three main objectives of electoral delimitation

The US Constitution and Supreme Court rulings have clearly mentioned three key requirements to be met by any process proposing an electoral delimitation scheme as shown in fig. 1: (i) proportional equality (i.e., approximately equal proportion of seat-to-population ration in all constituencies) (ii) spatial contiguity (which ensures cohesiveness of administration among territories) and (iii) compactness (which includes geographical, cultural and political closeness among voters and representatives) [2].

When the process of electoral delimitation results in constituencies that are unfairly disproportionate or unjustly contradict the principle of fair elections in cases of malapportionment or gerrymandering, the delimitation process must be scrutinised and revaluated, as shown by Schuster et al. in [4].

The use of automated districting procedures has the potential to ensure that systematic distortions of electoral outcomes are kept to a minimum. The denial of manipulating the primitive manual process of drawing out constituencies by political parties can be very useful to provide fair district maps [5].

In this paper we present a computational model for the process of electoral delimitation based on the quadtree data structure to automate the districting process applied in the demarcation of constituencies and further prove its efficiency to hold advantages over the existing methods for apportionment.

## II. LITERATURE SURVEY AND RELATED WORK

Research into political districting and electoral delimitation has been carried out from the view of a number of domains, including political, social and computational viewpoints [5]. A computational model, rather than traditional techniques suggested by Ricca F. et al. in [6] for political redistricting seems the most plausible as it has great power to prevent a number of delimitation problems including gerrymandering.

The multi kernel growth models for electoral delimitation were first suggested in the 1960s (Vickrey, 1961) [7] but lost their way after the 90s as no further progress was likely using this technique. The mathematical modeling efforts provided by Garfinkel and Nemhauser (1961) suggested an analogy to the common warehouse location problem, yet their approach failed to take into account the principal requirement of contiguity in the case of constituency demarcation [8].

Exact approaches which included techniques of continuity graphs by Nemoto and Hotta (2003) [9] and quadratic models by Li et al. (2007) [10] discuss methodologies for correct formulation of the problem of delimitation on smaller regions. However, as seen in real-world applications, the territory under study may have a much larger size and so the exact approach often fails.

Computational models often suggest an idealised plan for electoral delimitation, yet they likely fail to satisfy the fundamental requirement of contiguity. This led to many models having a post-processing phase in which contiguity is forced, often manually. Others tried to avoid this trouble by preliminary subdividing the elementary territorial units into smaller entities. Both approaches showed scope for improvement in the computational process used for delimitation [5].

The algorithms used for electoral delimitation in the United States as given by the U.S. Census Bureau [11], namely, (i) The Jefferson Method, (ii) The Hamilton Method, (iii) The Webster Method and (iv) The Huntington-Hill Method, provide only a method to determine the number of electoral seats, i.e., the number of constituencies in existing states, but remain silent about their actual shape and location within the state. Therefore, additional computation needs to be done in order to visualize the delimitation in a graphical way.

Moreover, traditional delimitation algorithms often suffered from the problem of "Gerrymandering", which refers to the practice of (re)drawing electoral district boundaries to advance the interests of the controlling political faction. This term was is a combination of the words- salamander and Gerry, the name of the governor of Massachusetts at the time, after he approved a district in the shape of a salamander to advance his own political interests. [12]

In summary, almost all of the computational models suggested fall short of satisfying all the objective criteria required for the delimitation exercise. This paper aims to resolve these problems associated with political districting (Verma, A.K., 2002) which include (i) selection of a correct and satisfactory method for the allocation of seats among several states (ii) redrawing the electoral boundaries of the constituencies in such a manner that the exercise does not fall into the trap of "gerrymandering" (iii) balancing the population of the constituencies in such a manner that it synchronizes with the legal and constitutional dictates for delimitation, using the quadtree data structure.

## III. TERMINOLOGY AND DATA STRUCTURES

*Quadtrees*: A quadtree is a non-linear data structure whose nodes are either leaves or have four children themselves [13]. In terms of applications, the term quadtree implies a class of representing geometric entities in a space of two dimensions that recursively decompose the space containing these entities into smaller blocks until the data in each block satisfy some condition (with respect, for example, to the block size, the number of block entities, the characteristics of the block entities, etc.) [14].

More formally, the term quadtree refers to a special tree data structure in which each internal node has four children. The entire space is initially represented in the root of the tree. The root node branches off into four children each representing a subregion of the parent. The union of all the siblings' subregions constitutes the entirety of the parent's region [14].

The basic objective of using quadtrees in geographical representations has been represented well using appropriate figures by Hunter, Gregory and Steiglitz in their work on "Operations on images using quadtrees" [13], as shown in fig. 2.

*Quadtrees for delimitation*: A quadtree used for the process of electoral delimitation contains population data for a specific geographical region as its node contents. Based on objective population criteria, each node may be further subdivided into four child nodes. The $i^{th}$ child node is associated with the $i^{th}$ quadrant of its parent's geographical region. This recursive process is continued till a population threshold can be met at every partition. These geographical partitions generated by the quadtree nodes can then be used to generate constituency demarcations in terms of quadrants on a map.

## IV. PROPOSED METHODOLOGY

### A. Quadtrees for geographical representation

Quadtrees are used to keep a balance between the precision and validity of the results obtained in geographical applications. A quadtree is useful to partition your search space into units of correlated spatial information as used for indexing in [15] and reduce a bigger problem to a set of problems of reduced size.

Quadtrees are classified according to the type of data they represent as point and area quadtrees [16]. The scope of this paper is based on area quadtrees.

Often, a quadtree is represented as a grid, with each square representing a node within the tree, as shown in fig. 3. This square in our application represents an electoral constituency generated using this data structure.

The subsequent image visualizes the method of subdividing nodes during a quadtree, starting with a "uni-root" node; in our case the whole area selected for delimitation.

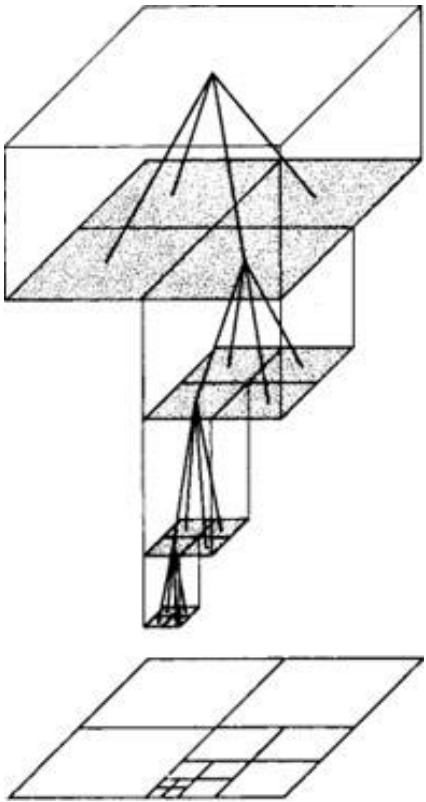

Fig. 2. Quadtrees for geographical area representation. Adapted from [13]

*B. Electoral delimitation using quadtrees*

The quadtree based approach to delimitation starts with the original image as a single one constituency and subdivides it further if a need to do so is felt. More specifically if the constituency in hand has population more than what is expected, it is subdivided into 4 smaller constituencies. Each constituency is associated to a node in the quadtree which makes it easy to store additional information if required any like constituency name, number, list of contesting candidates etc.

Finally, the leaves of the quadtree represent the actual plotted constituencies for the given map. Put in other words the count of the leaves gives us the total number of constituencies for the given population map of the concerned state or country and their arrangement within the map provides an outline of the constituency demarcation done for that particular region.

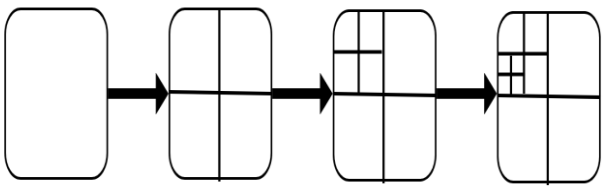

Fig. 3. Quadtrees as a grid

## V. ALGORITHM

1. Begin.
2. Construct a quadtree with a single node initially, which represents the entire image. This may be called the root node (uni-root).
3. Initialize current with root.
4. Initialize an empty map data structure called mp to map leaf nodes to their parent nodes.
5. Calculate 'n' as the number of dots in the part of the image associated with the current.
6. The required population 'p' is then x times n.
7. If 'p' exceeds Th:
   7.1 Logically divide the image under consideration into 4 quadrants, each represented by a node structure.
   7.2 Make current as the parent for these newly created nodes.
   7.3 Repeat steps 5. to 8. for each of the 4 children of current.
8. Else make mp[previous node] = current, since current is a leaf node .
9. After processing the entire map, merge the leaf nodes belonging to the same parent in case their population taken together falls below the threshold.
10. For every parent in map
    10.1.1. For every pair of children x, y that can be merged
        10.1.1.1 Let the region represented by x and y collectively be w.
        10.1.1.2 Add w to mp[parent]
        10.1.1.3 Remove x and y from mp[parent]
11. Stop

## VI. PSEUDOCODE

The quadtree based method proposed in this paper comprises of two major algorithms.

**1] Algorithm processRegion:** This algorithm processes the region associated with the passed node.

Algorithm processRegion (prev, n, countOfConstituencies, threshold, param, map)
{
// Inputs:
// 1. prev = The parent node of n.
// 2. n = node under consideration
// 3. countOfConstituencies = keeps track of the final constituencies' count.
// 4. threshold = maximum population, the region represented by n can have.
// 5. param = population per dot
// 6. map = contains a mapping between the parent node and child nodes only when the child nodes are leaves.

//Output:
// 1. countOfConstituencies updated value gives the total seats for the given population distribution and population to seat ratio.

dotcnt := countDots(n);
// countDots returns the count of dots in the region represented by n.

population := param * dotcnt;
  if population ⩽ threshold
{
        countOfConstituencies++;
        Add n to map[prev];

```
        // a valid constituency, hence no need to subdivide
          further.
        // Note prev as the parent of n.
        exit algorithm
  }
// Let q1, q2, q3 and q4 represent the 4 quadrant regions in
which the current region gets divided into.

 leftTop := Node(q1);
 rightTop :=Node(q2);
 leftBottom := Node(q3);
 rightBottom := Node(q4);
 n.children := [leftTop,rightTop,leftBottom,rightBottom];

//This makes n as the parent node for the newly created
child nodes.
//Now we need to process each of the four child nodes.

for child in n.children
 {
        CallprocessRegion(n, child,
          countOfConstituencies, threshold, param, map);
 }
}end Algorithm.
```

2] **Algorithm DelimitMap**: This algorithm is responsible for taking input, carrying out the delimitation activity on the given map and presenting the final output to the user.

```
Algorithm DelimitMap (map, threshold, x)
{
//Inputs:
// 1. map = image representing input region to be delimited
// 2. threshold = maximum population, a constituency can
have, alternatively population to seat ratio.
// 3. x = population each dot on the map represents.

//Output:
// 1. map containing constituency delimitation boundaries
marked on it.
// 2. countOfConstituencies gives the total seats for the
given population distribution and population to seat ratio.

RootNode := Node ( map );

// This creates a node representing the entire map as one
single region.

QuadTree := QuadTree (RootNode);
// Initialize a QuadTree with RootNode as its root
    countOfConstituencies := 0;

map <Parent,Children> map;
// Associative data structure to map leaf nodes to their parent
nodes.
// Parent represents parent node;
// Children contains nodes which are both leaf nodes and
children of Parent.
// Used for merging leaf nodes of the same parent if their
population put together falls below or equal to the threshold.

Call processRegion(NULL, RootNode,
countOfConstituencies, threshold, x, map);

for every parent in the map
{
  for every possible pair of children <x,y> in
  map[parent]
  {
    if param * (countDots(x)+countDots(y)) ⩽ threshold
    {
      MergedChild := Node(w);
      // implies x and y can be merged. Let the region
      represented by x and y be collectively represented
      by region w

      parent.children.remove(x);
      parent.children.remove(y);
      parent.children.add(MergedChild)
      //replace x and y with a single node MergedChild.
      countOfConstituencies–;
      }
   }
}
}
end Algorithm.
```

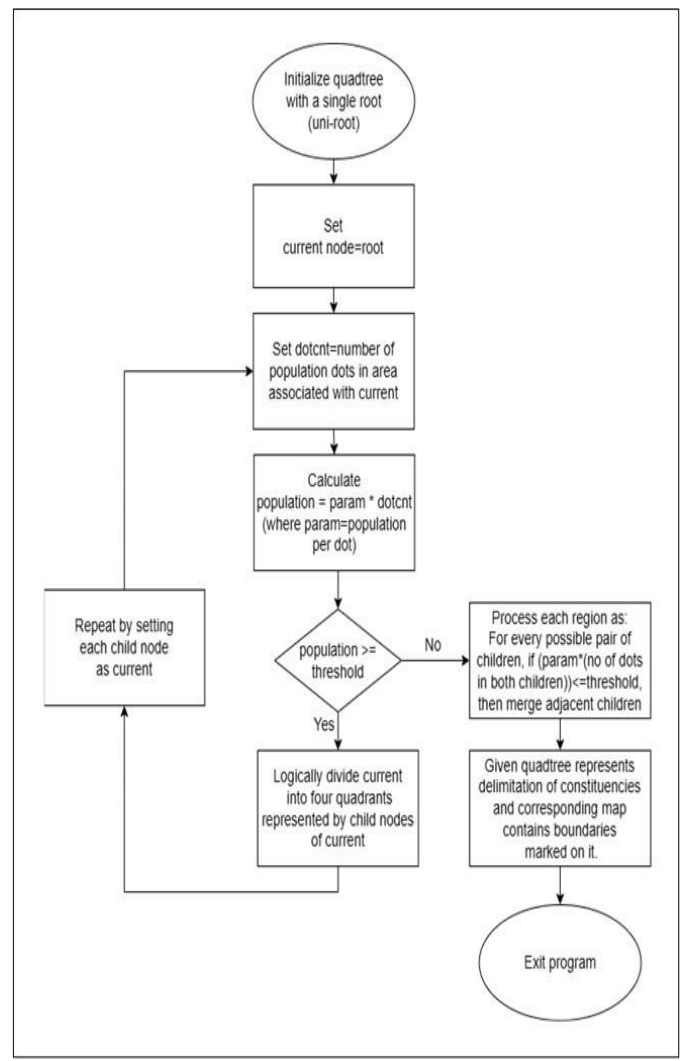

Fig. 4. Flowchart representing electoral delimitation process using quadtrees

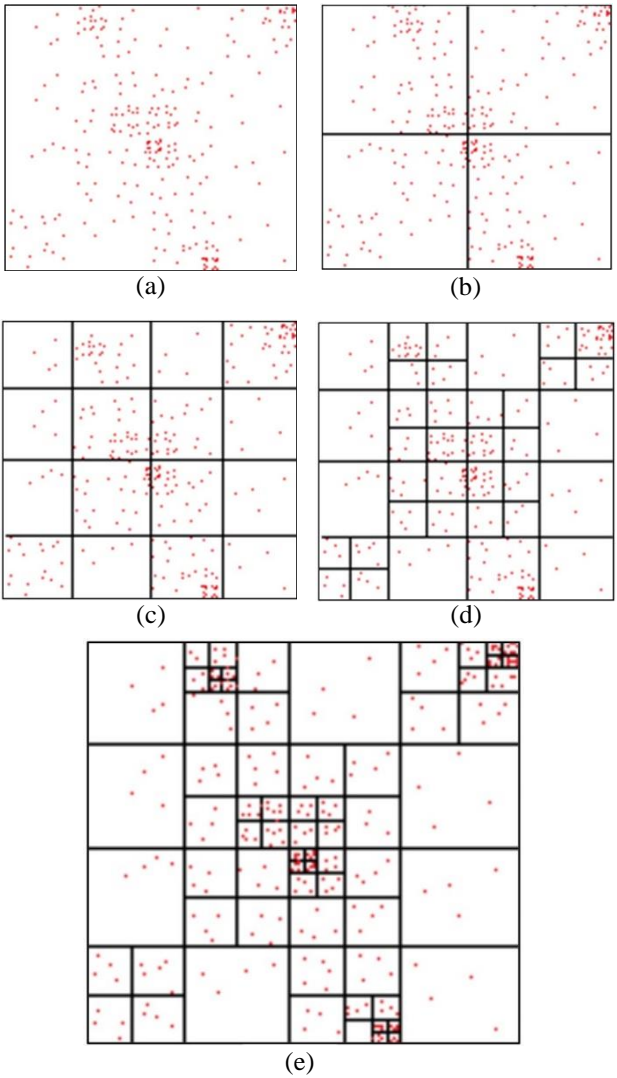

(a)  (b)  (c)  (d)  (e)

Fig. 5. Electoral delimitation using quadtrees (a) An image with population represented as dots having a predefined value (b) Initial division of area into four geographical quadrants. (c) Subsequent divisions into smaller geographic regions. (d) Further divisions into smaller geographic regions. (e) Final result of delimitation

## VII. RESULT DISCUSSION

The success of the quadtree based method shown by fig. 4 and fig. 5 for electoral delimitation can be proved with the help of a comparative study of results on a sample dataset as follows.

### A. Model followed by traditional techniques

The population data of four states (A, B, C and D) measured by the census department is used as a sample [11] given in table 1.

TABLE I. SAMPLE DATASET FOR ELECTORAL DELIMITATION

| State | Population |
|---|---|
| A | 2560 |
| B | 3315 |
| C | 995 |
| D | 5012 |

The U.S. Census Bureau has listed several methods (found on the official website) for apportionment of seats in states, each of which follows a predetermined mathematical approach which simply distributes the allotted number of seats, decided before, among the states. These traditional methods of delimitation will allocate a total of 20 seats for the given dataset producing the results given in tables 2-5 [11].

TABLE II. RESULTS OF THE HAMILTON/VINTON METHOD

| State | Population | Seats Apportioned |
|---|---|---|
| A | 2560 | 4 |
| B | 3315 | 5 |
| C | 995 | 2 |
| D | 5012 | 8 |

TABLE III. RESULTS OF THE WEBSTER METHOD

| State | Population | Seats Apportioned |
|---|---|---|
| A | 2560 | 4 |
| B | 3315 | 6 |
| C | 995 | 2 |
| D | 5012 | 8 |

TABLE IV. RESULTS OF THE JEFFERSON METHOD

| State | Population | Seats Apportioned |
|---|---|---|
| A | 2560 | 4 |
| B | 3315 | 6 |
| C | 995 | 1 |
| D | 5012 | 9 |

TABLE V. RESULTS OF THE HUNTINGTON-HILL METHOD

| State | Population | Seats Apportioned |
|---|---|---|
| A | 2560 | 4 |
| B | 3315 | 6 |
| C | 995 | 2 |
| D | 5012 | 8 |

### B. Mathematical model followed by quadtree based approach

The quadtree based electoral delimitation model suggested in this paper for the same dataset with a random distribution of population provides a visual representation of the demarcated constituencies as well. Thus, along with an idea of the total number of seats to be allocated for fair representation, it emphasises a bottom-up approach to the process of electoral delimitation.

*Formulation:* Given population data of four states (A, B, C and D), generate constituency boundaries such that they satisfy objectives of electoral delimitation.

*Mathematical Model:*
  1. Let '$x_i$' represent the population of the ith state
  2. If population '$x_i$'>threshold, divide state into 4 parts and generate constituencies with populations [$x_{i1}$, $x_{i2}$, $x_{i3}$, $x_{i4}$] and mark boundaries on the state map
  3. For each constituency 'k' with population '$x_{ik}$', recursively apply step 2 using '$x_{ik}$' as xi till each $x_{ik} \leqslant$ threshold
  4. For every possible pair of constituencies with populations [$x_{ij}$, $x_{ik}$] in a state, replace $x_{ij}$ and $x_{ik}$ with a single

constituency $x_{im}$ such that $x_{im} = x_{ij} + x_{ik}$ and redraw the merged constituency boundaries

5. Apply the model given by steps 1 to 4 to each state

*Results Generated:*

Along with an idea of the total number of seats to be allocated for fair representation, the model followed by the quadtree-based approach emphasises a bottom-up approach to the process of electoral delimitation. The results generated are represented in figure 6 and table 6.

TABLE VI. RESULTS OF THE QUADTREE-BASED METHOD

| State | Population | Seats Apportioned |
|---|---|---|
| A | 2560 | 4 |
| B | 3315 | 5 |
| C | 995 | 2 |
| D | 5012 | 8 |

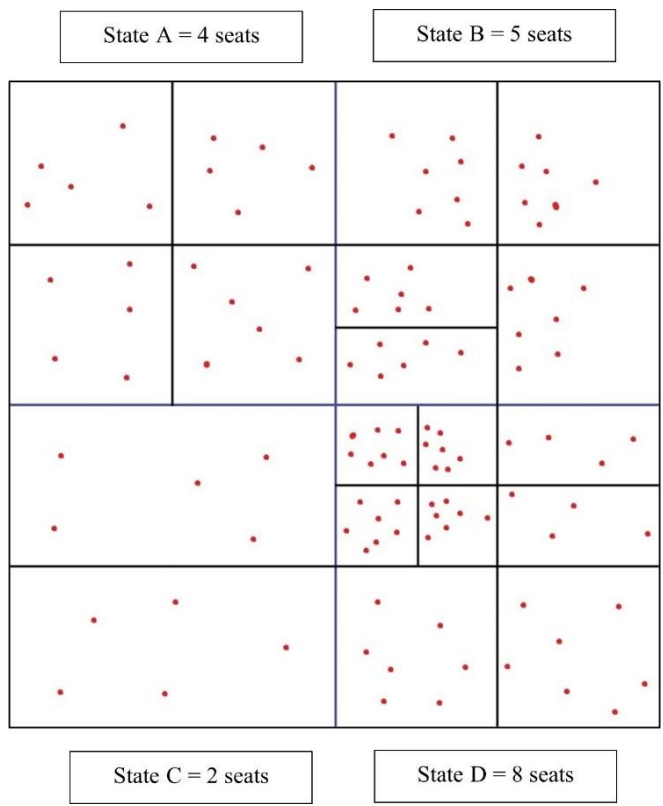

Fig. 6. Results of the quadtree based method of electoral delimitation with state boundaries in blue and constituency boundaries in black (1 dot=500 people)

Thus, the advantages held by the proposed methodology over traditional approaches can be considered and analysed from the following perspectives:

1. *Provides visual representation of allocated seats and boundaries*: Traditional methods of apportionment provide only an idea of how many seats to allocate in each state and fail to give a representation of the same on the map as they follow a purely mathematical approach. The quadtree based method instead gives a clear output of the delimitation process in the form of boundaries on a map.

2. *Reduced complexity of operations:* The proposed quad tree-based approach captures the delimitation information in a non-linear tree data structure, which makes further operations on the data cheaper in terms of complexity.

For instance, the process of searching for the constituency which contains a particular location is greatly simplified. The coordinates of the region represented by each node, starting with the root; can be compared with the coordinates of the location to be searched for. This, in the average case, would take time proportional to $\theta(\log_4(\text{input-size}))$. Thus, this method is significantly faster than iterating through all the constituencies individually, which would be the brute force approach to the above search taking $\theta(\text{input-size})$ time complexity.

Similarly, other operations like insert and delete also have asymptotic time complexity proportional to $\theta(\log_4(\text{input-size}))$ in the quadtree-based method, making it significantly faster.

VIII. CONCLUSION

The study of the problem of electoral delimitation and algorithms has been a matter of great interest for researchers over the years, driven by the challenge of proposing an efficient and automatable solution to support decision-makers in the process of demarcating constituencies.

The paper presented a quadtree data structure-based approach which can be a potential solution to a problem that has not yet been efficiently solved. This is because the quadtree based approach for electoral delimitation can tackle the majority of the problems associated with delimitation. Fairness in the delimitation process can be guaranteed as equality in representation can be achieved, a factor essential to democracy. Following a recursive automated algorithm, the constituencies generated in this process are uniform in shape, thus preventing political parties from tampering with the process by the process of gerrymandering using either of the techniques of cracking or packing. This proposed approach is also able to generate constituencies in a continuous pattern, sharing boundaries, thus attaining contiguity in delimitation.

Along with satisfying the constraints of political redistricting, the performance of the proposed algorithm has been analysed and evaluated against existing methods and proved to be a more efficient solution in terms of algorithmic complexity and boundary visualisation.

Thus, the solution proposed using quadtrees is theoretically able to solve all problems of electoral delimitation proposed above and is likely to be effective in its implementation.

IX. FUTURE SCOPE

The electoral delimitation carried out using the quadtree approach has been conducted on theoretical data to provide a basis for proving the efficiency of this approach in contrast to the traditional methods used for delimitation.

The districting process is an important issue not just in electoral affairs but also in business and industry. The constituency boundaries developed in this paper can also be applied to determine a variety of other territorial partitions including: (i) school districts, (ii) geo-marketing districts

(iii) postal districting maps, and any other application, where population data can be used for partitioning.

However, implementation problems associated with the application of this method remain a subject of future interest as even though the principle used for partitioning is scientifically acceptable, the acceptance of the allocated boundaries in a social sense among the public remains unpredictable.

Thus, all in all, considering that electoral delimitation not only plays a significant role in politics but also representation of society, certain case-based modifications to the proposed algorithm can go a long way in achieving socially as well as politically acceptable electoral constituency boundaries.